\documentclass[sigconf,10pt]{acmart}

\usepackage{booktabs} 
\usepackage{CJKutf8}
\usepackage{latexsym}
\usepackage{makecell}
\usepackage{epsfig}
\usepackage{algorithm, algpseudocode}
\usepackage{booktabs}
\usepackage{multirow}
\usepackage{color}
\usepackage{subfigure}
\usepackage{url}
\usepackage{mathrsfs}
\usepackage{amsmath}
\usepackage{bm}
\usepackage{graphicx}

\newcommand{\mlp}{\operatornamewithlimits{MLP}}

\newcommand{\cnn}{\operatornamewithlimits{CNN}}
\newcommand{\maxpool}{\operatornamewithlimits{MaxPooling}}
\newcommand{\selfatt}{\operatornamewithlimits{SelfAtt}}
\newcommand{\docvec}{\operatornamewithlimits{Doc2vec}}

\newcommand{\secref}[1]{Section \ref{#1}}
\newcommand{\figref}[1]{Figure \ref{#1}}

\newcommand{\algoref}[1]{Algorithm \ref{#1}}
\newcommand{\tabref}[1]{Table \ref{#1}}
\newcommand{\bi}[1]{\textbf{\textit{#1}}}

\copyrightyear{2020}
\acmYear{2020}
\setcopyright{acmcopyright}
\acmConference[SIGMOD'20]{Proceedings of the 2020 ACM SIGMOD International Conference on Management of Data}{June 14--19, 2020}{Portland, OR, USA}
\acmBooktitle{Proceedings of the 2020 ACM SIGMOD International Conference on Management of Data (SIGMOD'20), June 14--19, 2020, Portland, OR, USA}
\acmPrice{15.00}
\acmDOI{10.1145/3318464.3386132}
\acmISBN{978-1-4503-6735-6/20/06}

\begin{document}
	
\fancyhead{}

\title{\textit{AliCoCo}: Alibaba E-commerce Cognitive Concept Net 
}

%
%
%
%
%
%
%

\author{Xusheng Luo}
\authornote{Corresponding author}
\affiliation{
	\institution{Alibaba Group, Hangzhou, China}
}
\email{lxs140564@alibaba-inc.com}

\author{Luxin Liu, Le Bo, Yuanpeng Cao, Jinhang Wu, Qiang Li}
\affiliation{
	\institution{Alibaba Group, Hangzhou, China}
}
\author{Yonghua Yang, Keping Yang}
\affiliation{
	\institution{Alibaba Group, Hangzhou, China}
}
\author{Kenny Q. Zhu}
\authornote{Kenny Q. Zhu was partially supported by NSFC grant 91646205 and Alibaba Visiting Scholar Program.}
\affiliation{
	\institution{Shanghai Jiao Tong University, Shanghai, China}
}


%
%

\copyrightyear{2020} 
\acmYear{2020} 
\acmConference[SIGMOD '20]{ACM SIGMOD/PODS International Conference on Management of Data}{June 14--19, 2020}{Portland, OR, USA}
\acmBooktitle{2020 International Conference on Management of Data (SIGMOD '20), June 14--19, 2020, Portland, OR, USA}
\acmPrice{15.00}
\acmDOI{10.1145/3357384.3357812}
\acmISBN{978-1-4503-6976-3/19/11}

\begin{abstract}
One of the ultimate goals of e-commerce platforms is to satisfy various shopping needs for their customers.
Much efforts are devoted to creating taxonomies or ontologies in e-commerce towards this goal.
However, user needs in e-commerce are still not well defined, and none of the existing ontologies has the enough depth and breadth for universal user needs understanding.
The semantic gap in-between
prevents shopping experience from being more intelligent.
In this paper, 
we propose to construct a large-scale e-commerce \textbf{Co}gnitive \textbf{Co}ncept net named ``\textbf{AliCoCo}'', which is practiced in \textbf{Ali}baba, the largest Chinese e-commerce platform in the world.
We formally define user needs in e-commerce,
then conceptualize them as nodes in the net.
We present details on how AliCoCo is constructed semi-automatically and its successful, ongoing and potential applications in e-commerce. 
\end{abstract}



\maketitle
\begin{CJK}{UTF8}{gbsn}
\section{Introduction}
\label{sec:intro}


One major functionality of e-commerce platforms is to match the shopping need of a customer to a small set of items from an enormous candidate set.
With the rapid developments of search engine and recommender system,
customers are able to quickly find those items they need.
However, the experience is still far from ``intelligent''.
One significant reason is that there exists a huge semantic gap between what users need in their mind and how the items are organized in e-commerce platforms.
The taxonomy to organize items in Alibaba (actually almost every e-commerce platforms) is generally based on \textbf{CPV} (Category-Property-Value):
thousands of categories form a hierarchical structure according to different granularity, and properties such as color and size are defined upon each leaf node.
It is a natural way of organizing and managing billions of items in nowadays e-commerce platform, and already becomes the essential component in downstream applications including search and recommendation.
However, existing taxonomies or ontologies in e-commerce are difficult to interpret various user needs comprehensively and accurately due to the semantic gap, which will be explained in the following two scenarios.

For years, e-commerce search engine is teaching our users how to input keywords wisely so that the wanted items can be quickly found. 
However, it seems keyword based searching only works for those users who know the exact product they want to buy. 
The problem is, users do not always know the exact product. 
More likely what they have in mind is a type or a category of products, with some extra features. 
Even worse, they only have a scenario or a problem but no idea what items could help. 
In these cases, a customer may choose to conduct some research outside the e-commerce platform to narrow down to an exact product, which harms the user experience and making e-commerce search engine not intelligent at all. 
If tracing back to the source, 
the real reason behind this is that existing ontologies in e-commerce doesn't contain structured knowledge indicating what products are needed for an ``outdoor barbecue'' 
or what is ``preventing the olds from getting lost''. 
Typing search queries like these inevitable leads to user needs mismatch and query understanding simply degenerates to key words matching.

\begin{figure*}[th]
	\centering
	\epsfig{file=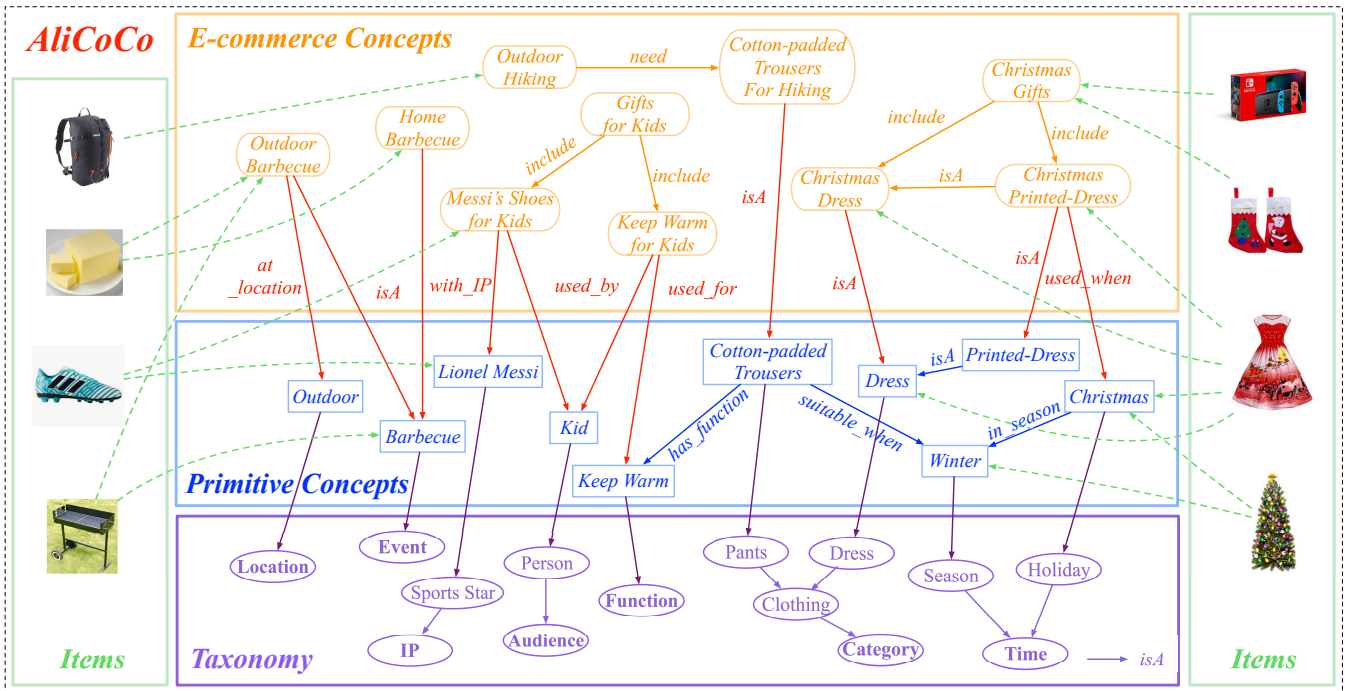, width=2.1\columnwidth}
	\caption{Overview of ``AliCoCo'', which consists of four layers: e-commerce concepts, primitive concepts, taxonomy and items.}
	\label{fig:overview}
\end{figure*}

The same problem exists in item recommendation. 
Due to the prohibitive size of transaction data in real-world industry scenario, recommendation algorithms widely adopt the 
idea of item-based CF \cite{sarwar2001item}, which
can recommend from very large set of options with relatively small amount of computation, 
depending on the pre-calculated similarity between item pairs.
The recommender system uses user's historical behaviors as triggers to recall a small set of most similar items as candidates, 
then recommends items with highest weights after scoring with a ranking model.
A critical shortcoming of this framework is that it is not driven by user needs in the first place, 
which inevitably leads to a dilemma where items recommended are hard to be explained 
except for trivial reasons such as ``similar to those items you have already viewed or purchased''.
Besides, it also prevents the recommender system from jumping out of 
historical behaviors to explore other implicit or latent user interests.
Therefore, despite the widespread of its use,
the performance of current recommendation systems is still under criticism. 
Users are complaining that some recommendation results are redundant and lack novelty, 
since current recommender systems can only satisfy very limited user needs such as the needs for a particular category or brand.
The lack of intermediate nodes in current e-commerce ontologies that can represent various user needs constrains the development of recommender systems.

In this paper, we attempt to bridge the semantic gap between actual user needs and existing ontologies in e-commerce platforms 
by building a new ontology towards universal user needs understanding.
It is believed that the cognitive system of human beings is based on \textit{concepts} \cite{murphy2002thebig,bloom2003glue},
and the taxonomy and ontology of concepts give humans the ability to understand \cite{wu2012probase}.
Inspired by it, we construct the ontology mainly based on concepts and name it ``\textbf{AliCoCo}'': \textbf{Co}gnitive \textbf{Co}ncept Net in \textbf{Ali}baba.
Different from most existing e-commerce ontologies, 
which only contain nodes such as categories or brands, 
a new type of node, e.g.,
``Outdoor Barbecue'' and ``Keep Warm for Kids'', is introduced as 
bridging concepts connecting user and items to satisfy some 
high-level user needs or shopping scenarios. 
Shown in the top of \figref{fig:overview}, we call these nodes ``\textbf{e-commerce concepts}'',
whose structure represents a set of items from different categories 
with certain constraints (more details in \secref{sec:ecommerce}) .
For example, ``Outdoor Barbecue'' is one such e-commerce concept,  
consisting of products such as grills, butter and so on, 
which are necessary items to host a successful outdoor barbecue party.
Therefore, AliCoCo is able to help search engine directly suggest a customer 
``items you will need for outdoor barbecue'' 
after he inputs keyword ``barbecue outdoor'',
or help recommender system remind him of preparing things that can ``keep warm for your kids'' as 
there will be a snowstorm coming next week.

\begin{figure}[th]
	\centering
	\epsfig{file=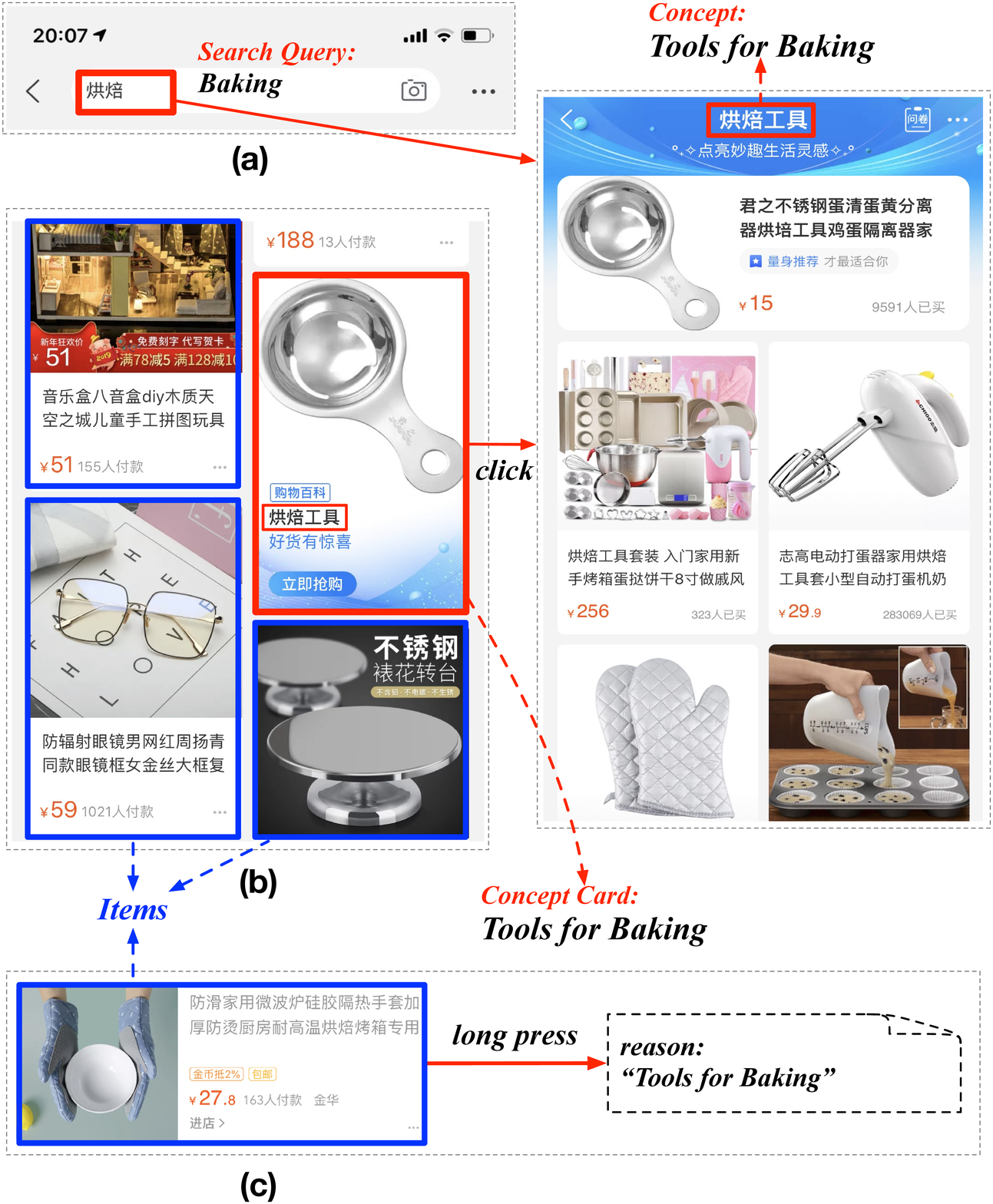, width=\columnwidth}
	\caption{Three real examples of user-needs driven e-commerce.
		(a): Queries trigger concept cards in semantic search. (b): Display concepts directly to users as cards with a set of related items. (c): Concepts act as explanations in search and recommendation.
	}
	\label{fig:cloud}
\end{figure}

There are several possible practical scenarios in which 
applying such e-commerce concepts can be useful.
The first and most natural scenario is directly displaying those concepts to users 
together with its associated items.
\figref{fig:cloud}(a/b) shows the real implementation of this idea in 
\textit{Taobao} \footnote{\url{http://www.taobao.com}} App.
Once a user typing ``Baking'' (a), 
he will enter into a page (right) where different items for baking are displayed, making the search experience a bit more intelligent.
It can also be integrated into recommender systems.
Among normal recommended items, 
concept ``Tools for Baking'' is displayed to users as a card with its name and the picture of a representative item (b).
Once a user clicks on it, he will enter into the page on the right.
In this way, the recommender system is acting like a salesperson in a shopping mall, 
who tries to guess the needs of his customer and and then suggests how to satisfy
them. 
If their needs are correctly inferred, users are more likely to accept 
the recommended items.
Other scenarios can be providing explanations in search or recommendation as 
shown in \figref{fig:cloud}(c).
While explainable recommendation attracts much research attention 
recently \cite{zhang2018explainable}, 
most existing works are not practical enough for industry systems,
since they are either too complicated 
(based on NLG \cite{zanker2010knowledgeable,cleger2012explaining}), or too trivial 
(e.g., ``how many people also viewed'' \cite{costa2018automatic,li2017neural}).
Our proposed concepts, on the contrary, precisely conceptualize user needs and are
easy to understand.

\begin{itemize}
	\item We claim that current ontologies in e-commerce platforms are unable to represent and understand actual user needs well and therefore prevent shopping experience from being more intelligent. To bridge the semantic gap in between, we formally define user needs in e-commerce and propose to build an end-to-end large comprehensive knowledge graph called ``AliCoCo'', where the ``concept'' nodes can explicitly represent various shopping needs for users.
	\item To construct such a large-scale knowledge graph, we adopt a semi-automatic way by combining both machine learning efforts and manual efforts together. We detailed introduce the four-layer structure of AliCoCo and five non-trivial technical components. For each component, we formulate the problem, point out the challenge, describe effective solutions and give thorough evaluations.
	\item AliCoCo is already gone into production in Alibaba, the largest e-commerce platform in China. It benefits a series of applications including search and recommendation.
	We believe the idea of user needs understanding can be further applied in more e-commerce productions. There is ample room for imagination and further innovation in ``user-needs driven'' e-commerce.
\end{itemize}

The rest of paper is organized as follows:
First we give an overview
of AliCoCo (\secref{sec:overview}), then present how we construct each of the four layers:
Taxonomy (\secref{sec:taxonomy}), Primitive Concepts (\secref{sec:primitive}), E-commerce Concepts (\secref{sec:ecommerce}), and Item Associations (\secref{sec:item}).
\secref{sec:eval} shows overall statistics of AliCoCo and evaluations of five main technical modules.
Then we discuss some successful, ongoing and potential applications in \secref{sec:application}.
\secref{sec:related} mentions related works, and finally, \secref{sec:conclusion} gives a conclusion and delineates possible future work.

\section{Overview} 
\label{sec:overview}

AliCoCo provides an alternative to describing and understanding user needs and items in e-commerce within the same, universal framework.
As shown in \figref{fig:overview}, 
AliCoCo consists of four components:
\textbf{E-commerce Concepts}, \textbf{Primitive Concepts}, \textbf{Taxonomy} and \textbf{Items}.

As the core innovation, 
we represent various user needs as \textbf{E-commerce Concepts} (orange boxes) in the top layer of \figref{fig:overview}.
E-commerce concepts are short, coherent and plausible phrases such as ``outdoor barbecue'', ``Christmas gifts for grandpa''
or ``keep warm for kids'', which describe specific shopping scenarios.
User needs in e-commerce are not formally defined previously,
hierarchical categories and browse nodes \footnote{\url{https://www.browsenodes.com/}} are usually used to represent user needs or interests \cite{zhou2018deep}.
However, we believe user needs are far broader than categories or browse nodes. 
Imaging a user who is planning an outdoor barbecue, or who is concerned with how to get rid of a raccoon in his garden.
They have a situation or problem but do not know what products can help.
Therefore, user needs are represented by various concepts in AliCoCo,
and more details will be introduced in \secref{sec:ecommerce}.

To further understand high-level user needs (aka. e-commerce concepts), 
we need a fundamental language to describe each concept.
For example, ``outdoor barbecue'' can be expressed as ``\textit{<Event: Barbecue>} | \textit{<Location: Outdoor>} | \textit{<Weather: Sunny>} | ...''.
Therefore, 
we build a layer of \textbf{Primitive Concepts}, where
``primitive'' means concept phrases in this layer are relatively short and simple such as ``barbecue'', ``outdoor'' and ``sunny'' (blue boxes in \figref{fig:overview}), 
comparing to e-commerce concepts above which are compound phrases in most cases.
To categorize all primitive concepts into classes, a \textbf{Taxonomy} in e-commerce is also defined, where classes with different granularities form a hierarchy via \textit{isA} relations.
For instance, there is a path top-down being ``\textit{Category->ClothingAndAccessory->Clothing->Dress}'' in the taxonomy (purple ovals in \figref{fig:overview}) .

We also define a schema on the taxonomy, 
to describe relations among different primitive concepts. For example, there is a relation ``\textit{suitable\_when}'' defined between ``class:\textit{ Category-Clothing->Pants}'' and ``class: \textit{Time->Season}'', so the primitive concept ``cotton-padded trousers'' is ``suitable\_when'' the season is ``winter''. 

In the layer of \textbf{Items}, billions of items \footnote{Items are the smallest selling units on Alibaba. Two iPhone Xs Max (each of them is an item) in two shops have different IDs.} on Alibaba are related with both primitive concepts and e-commerce concepts.
Primitive concepts are more like the properties of 
items, such as the color or the size.
However, the relatedness between e-commerce concepts and items represents that certain items are necessary or suggested under a particular shopping scenario. 
As the example shown in \figref{fig:overview},
items such as grills and butter are related to the e-commerce concept ``outdoor barbecue'', 
while they can not be associated with the primitive concept ``outdoor'' alone.

Overall, we represent user needs as e-commerce concepts, then adopt primitive concepts with a class taxonomy to describe and understand both user needs and items in the same framework. 
Besides, e-commerce concepts are also associated directly with items, to form the complete structure of AliCoCo.

\section{Taxonomy}
\label{sec:taxonomy}

\begin{figure}[th]
	\centering
	\epsfig{file=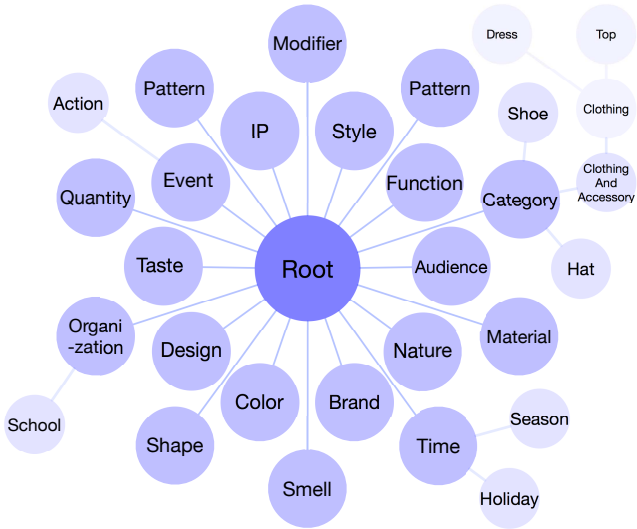, width=\columnwidth}
	\caption{Overview of the taxonomy in AliCoCo.}
	\label{fig:primitive}
\end{figure}

The taxonomy of AliCoCo is a hierarchy of pre-defined classes
to index million of (primitive) concepts.
A snapshot of the taxonomy is shown in \figref{fig:primitive}.
Great efforts from several domain experts are devoted to manually define the whole taxonomy.
There are $20$ classes defined in the first hierarchy, 
among which the following classes are specially designed for e-commerce, including ``\textit{Category}'', ``\textit{Brand}'', ``\textit{Color}'', ``\textit{Design}'',
``\textit{Function}'', ``\textit{Material}'',
``\textit{Pattern}'', ``\textit{Shape}''
``\textit{Smell}'', ``\textit{Taste}'' and 
``\textit{Style}'',
where the largest one is ``\textit{Category}'' having nearly $800$ leaf classes, since the categorization of items is the backbone of almost every e-commerce platform.
Other classes such as ``\textit{Time}'' and  ``\textit{Location}'' are more close to general-purpose domain.
One special class worth mentioning is ``\textit{IP}'' (Intellectual Property), 
which contains millions of real world entities such as famous persons, movies and songs.
Entities are also considered as primitive concepts in AliCoCo.
The $20$ classes defined in the first hierarchy of the taxonomy are also called ``domains''.

\section{Primitive Concepts}
\label{sec:primitive}
Primitive concepts with a class taxonomy are expected to describe every item and user need in e-commerce accurately and comprehensively.
They are the fundamental building blocks for understanding high-level shopping needs of our customers.
In this section, we mainly introduce how we mine these raw primitive concepts (can be seen as vocabulary) and then organize them into the hierarchical structure.

\subsection{Vocabulary Mining}
\label{sec:mining}
There are two ways of enlarging the size of primitive concepts once the taxonomy is defined.
The first one is to incorporate existing knowledge from multiple sources through ontology matching.
In practice, we mainly adopt rule-based matching algorithms, together with human efforts to manually align the taxonomy of each data source. Details will not be introduced in this paper.

The second one is to mine new concepts from large-scale text corpus generated in the domain of e-commerce such as search queries, product titles, user-written reviews and shopping guides.
Mining new concepts of specific classes can be formulated as \textit{sequence labeling} task,
where the input is a sequence of words and the output is a sequence of predefined labels.
However, the hierarchical structure of our taxonomy is too complicated for this task, 
so we only use the $20$ first-level classes as labels in practice.

\begin{figure}[th]
	\centering
	\epsfig{file=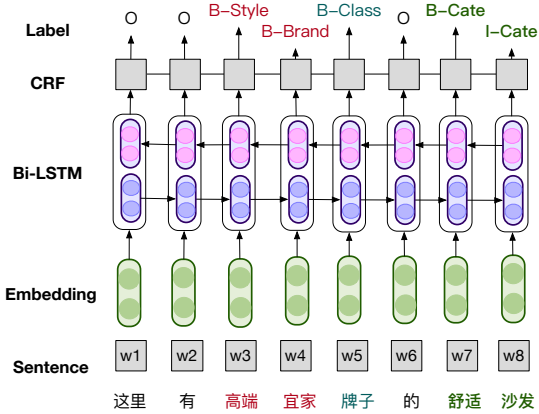, width=0.85\columnwidth}
	\caption{Principle architecture of a BiLSTM-CRF model
	}
	\label{fig:bilstm_crf}
\end{figure}

\figref{fig:bilstm_crf} shows the principle architecture of a BiLSTM-CRF model,
which is the state-of-the-art model for various sequence labeling tasks \cite{huang2015bidirectional,reimers2017optimal}.
BiLSTM-CRF model consists of a BiLSTM layer and a CRF layer, 
where BiLSTM (Bidirectional-LSTM) enables the
hidden states to capture both historical and future
context information of the words and 
CRF (Conditional Random Field) considers the correlations
between the current label and neighboring
labels.

All the automatically mined \textit{concept: class} pairs are then manually checked to ensure the correctness. 
Details will be introduced in \secref{sec:eval_mining}. 
Once the class is determined, a surface form then becomes a true primitive concept, and each concept will be assigned a unique ID.
There can be several primitive concepts with the same name but different IDs (meanings), 
giving AliCoCo the ability to disambiguate raw texts.

\subsection{Hypernym Discovery}
\label{sec:isa}
Once primitive concepts of $20$ first-level classes (domains) are mined,
we continue to classify each primitive concept into fine-grained classes within each domain.
In each domain, this task can be formulated as \textit{hypernym discovery}, 
where we have to predict the hyponym-hypernym relations between arbitrary pair of primitive concepts.
In practice, we exploit a combination of two methods: 
an unsupervised pattern-based method and a supervised projection learning model.

\subsubsection{Pattern based}
The pattern-based method for hypernym discovery was pioneered by Hearst \cite{hearst1992automatic}, who defined specific textual patterns like ``\textit{Y such as X}'' to mine hyponym-hypernym pairs from corpora.
This approach is known to suffer from low recall because it assumes that hyponym-hypernym pairs co-occur in one of these patterns, 
which is often not true when matching the patterns in corpora.
Besides those patterns,
we adopt other rules to directly discover hypernyms using some special grammar characteristics of Chinese language such as ``XX裤 (XX pants)'' must be a ``裤 (pants)'', etc.

\subsubsection{Projection learning}
The general idea of projection learning is to learn a function that takes as input the word embedding of a possible hyponym $p$ and a candidate hypernym $h$ and outputs the likelihood that there is a hypernymy relationship between $p$ and $h$. 
To discover hypernyms for a given hyponym $p$, we apply this decision function to all candidate hypernyms, and select the most likely ones.
Given a pair of candidate $p$ and $h$, we first obtain their 
word embeddings $\bi{p}$ and $\bi{h}$ through a lookup table where embeddings are pertained on e-commerce corpus. 
Then we use a projection tensor \bi{T} to measure how possible there is a hypernymy relation. 
In $k$th layer of \bi{T}, we calculate a score $s^k$ as:
\begin{equation}
s^k = \bi{p}^T\bi{T}^k\bi{h}
\end{equation}
where $\bi{T}^k$ is matrix and $k \in [1, K]$.
Combining $K$ scores, we obtain the similarity vector \bi{s}.
After apply a fully connected layer with sigmoid activation function,
we get the final probability $y$:
\begin{equation}
y = \sigma(\bi{W}\bi{s}+\bi{b})
\end{equation}

\subsubsection{Active learning}
Since labeling a large number of hyponym-hypernym pairs for each domain clearly does not scale, 
we adopt \textit{active learning} as a more guided approach to select examples to label so that we can economically learn an accurate model by reducing the annotation cost.
It is based on the premise that a model can get better performance if it is allowed to prepare its own training data, by choosing the most beneficial data points and querying their annotations from annotators.
We propose an uncertainty and high confidence sampling strategy (UCS) to select samples which can improve model effectively.
The iterative active learning algorithm is shown in \algoref{alg:al}.

\begin{algorithm}
	\small
	\caption{UCS active learning algorithm}
	\label{alg:al}
	\textbf{Input}: unlabeled dataset $D$, test dataset $T$, scoring function $f(\cdot,\cdot)$, human labeling $H$, 
	the number of human labeling samples in each iteration $K$;
	\textbf{Output}: scoring function $\hat f(\cdot,\cdot)$, predict score $S$ \ \ \ \ \ \ \ \ \ \  \ \
	\begin{algorithmic}[1]
		\Procedure{AL}{$D, D_0, T, f, H, K$}
		\State $i \gets 0$
		\State $D_0 \gets random\_select(D, K)$
		\State $L_0 \gets H(D_0)$
		\State $D \gets D - D_0$
		\State $\hat f, fs \gets train\_test(f, L_0, T)$
		\State $S \gets \hat f(D)$
		\Repeat
		\State $p_i = \frac{|S_i-0.5|}{0.5}$ 
		\State $D_{i+1} \gets D(Top(p_i, \alpha K)) \bigcup D(Bottom(p_i, (1-\alpha) K)) $ 
		\State $L_{i+1} \gets H(D_{i+1}) \bigcup L_i$
		\State $D \gets D - D_0$
		\State $\hat f, fs \gets train\_test(f, L_{i+1}, T)$
		\State $S \gets \hat f(D)$
		\Until{$fs$ not improves in n step} 
		\EndProcedure
	\end{algorithmic}
\end{algorithm}

As line 3 to 7 show, 
we first randomly select a dataset $D_0$ which contains $K$ samples from the unlabeled dataset $D$ and ask domain experts to label the samples from $D_0$.
As a result, we obtain the initial labeled dataset $L_0$ and $D_0$ is removed from the $D$. 
Then, we train the projection learning model $f$ using $L_0$ and test the performance on the test dataset $T$. $fs$ is the metrics on $T$. 
At last, we predict the unlabeled dataset $D$ using the trained $\hat f$ and get the score $S_0$.

Next, we iteratively select unlabeled samples to label and use them to enhance our model. 
We propose an active learning sampling strategy named uncertainty and high confidence sampling (UCS) which select unlabeled samples from two factors.
The first factor is based on classical uncertainty sampling (US)  \cite{lewis1994heterogeneous}.
If the prediction score of a sample is close to $0.5$, it means the current model is difficult to judge the label of this sample. 
If the expert labels this example, the model can enhance its ability by learning this sample. 
We calculate this probability by $\frac{|S_i-0.5|}{0.5}$ in line $9$.
Besides, we believe those samples with high confidence are also helpful in the task of hypernym discovery, since the model is likely to predict some difficult negative samples as positive with high confidence when encountering relations such as \textit{same\_as} or \textit{similar}.
The signal from human labeling can correct this problem in time.
Thus, we select those samples with high scores as well in line $10$.
In addition, we utilize $\alpha$ to control the weight of different sampling size.
Then, we get the new human labeled dataset which can be used to train a better model. 
As a result, with the number of labeled data increases, 
the performance of our model will also increase. 

Finally, this iterative process will be stopped when the performance of the model $fs$ does not increase in $n$ rounds. 
During the process, we not only get a better model but also reduce the cost of human labeling.

\section{E-commerce Concepts}
\label{sec:ecommerce}

In the layer of e-commerce concepts,
each node represents a specific shopping scenario,
which can be interpreted by at least one primitive concept.
In this section,
we first introduce the high criteria of a good e-commerce concept using several examples, 
then show how we generate all those e-commerce concepts  and further propose an algorithm to link e-commerce concepts to the layer of primitive concepts.

\subsection{Criteria}

As introduced in \secref{sec:overview},
user needs are conceptualized as e-commerce concepts in AliCoCo, and a good e-commerce concept should satisfy the following criteria:

\noindent
\textbf{(1) E-commerce Meaning.}
It should let anyone easily think of some items in the e-commerce platform, which means it should naturally represent a particular shopping need. Phrases like ``blue sky'' or ``hens lay eggs'' are not e-commerce concepts, 
because we can hardly think of any related items.

\noindent
\textbf{(2) Coherence.}
It should be a coherent phrase. Counter-examples can be 
``gift grandpa for Christmas'' or ``for kids keep warm'',
while the coherent ones should be ``Christmas gifts for grandpa'' and ``keep warm for kids''.

\noindent
\textbf{(3) Plausibility.}
It should be a plausible phrase  according to commonsense knowledge. Counter-examples can be ``sexy baby dress'' or 
``European Korean curtain'' since we humans will not describe a dress for babies using the word ``sexy'' and a curtain can not be in both European style and Korean style.

\noindent
\textbf{(4) Clarity.}
The meaning of an e-commerce concept should be clear and easy to understand.
Counter-examples can be ``supplementary food for children and infants'' where the subject of this can be either older-aged children or newborns. This may lead to a confusion for our customers.

\noindent
\textbf{(5) Correctness.}
It should have zero pronunciation or grammar error.

\subsection{Generation}

There is no previous work on defining such e-commerce concepts and few on mining such phrases from texts. 
In practice, we propose a two-stage framework: 
firstly we use two different ways to generate large amount of possible e-commerce concept candidates,
then a binary classification model is proposed to 
identify those concepts which satisfy our criteria.

\subsubsection{Candidate Generation}
There are two different ways to generate concept candidates. 
The first is mining raw concepts from texts.
In practice, we adopt AutoPhrase\cite{shang2018automated} to mine possible concept phrases from large corpora in e-commerce including search queries, product titles, user-written reviews and shopping guidance written by merchants.
Another alternative is to generating new candidates using existing primitive concepts. 
For example, we combine ``\textit{Location: Indoor}'' with ``\textit{Event: Barbecue}'' to get a new concept ``indoor barbecue'', which is not easy to be mined from texts since it's a bit unusual.
However, it is actually a quite good e-commerce concept 
since one goal of AliCoCo is to cover as many user needs as possible.
The rule to combine different classes of primitive concepts is using some automatically mined then manually crafted patterns.
For example, we can generate a possible concept ``warm hat for traveling'' using a pattern ``[\textit{class: Function}] [\textit{class: Category}] for [\textit{class: Event}]''.
\tabref{tab:pattern} shows some patterns used in practice and corresponding e-commerce concepts, including some bad ones waiting to be filtered out in the following step.
If a single primitive concept satisfies all five criteria, it can be regarded as an e-commerce concept as well. 

\begin{table*}[th]
	\centering
	\small
	\begin{tabular}{|l|c|c|}
		\hline
		\bi{Pattern} & \bi{Good Concept} & \bi{Bad Concept} \\
		\hline
		[\textit{class: Function}] [\textit{class: Category}] for [\textit{class: Event}] & warm hat for traveling & warm shoes for swimming \\
		\hline
		[\textit{class: Style}] [\textit{class: Time->Season}] [\textit{class: Category}] & British-style winter trench coat &
		casual summer coat \\
		\hline
		[\textit{class: Location}] [\textit{class: Event->Action}] [\textit{class: Category}] & British imported snacks &
		Bird's nest imported from Ghan \\
		\hline
		[\textit{class: Function}] for [\textit{class: Audience->Human}] & health care for olds &
		waterproofing for middle school students \\
		\hline
		[\textit{class: Event->Action}] in [\textit{class: Location}]  &  traveling in European &
		Bathing in the classroom \\
		\hline
	\end{tabular}
	\caption{Some patterns used to generate e-commerce concepts.}
	\label{tab:pattern}
\end{table*}

\subsubsection{Classification}
\label{sec:classification}
To automatically judge whether a candidate concept satisfies the criteria of being a qualified e-commerce concept or not, 
the main challenge is to test its plausibility.
For the other four criteria,
character-level and word-level language models
and some heuristic rules are able to meet the goal.
However, it is difficult for machines to grasp
commonsense knowledge as we humans do to 
know that ``sexy'' is not suitable to describe a dress when it's made for a child.
Moreover, the lack of surrounding contexts makes the problem more challenging, since our concepts are too short (2-3 words on average).

\begin{figure}[th]
	\centering
	\epsfig{file=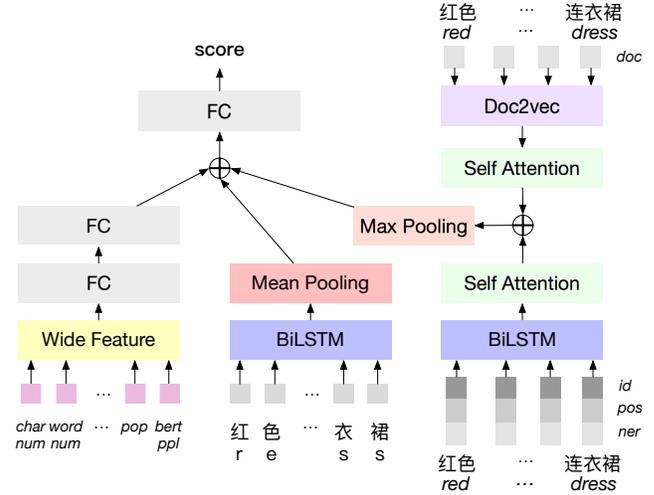, width=\columnwidth}
	\caption{Overview of knowledge-enhanced deep model for e-commerce concept classification.
	}
	\label{fig:classification}
\end{figure}

To tackle this problem, 
we propose a knowledge-enhanced deep classification model to first link each word of a concept to an external knowledge base then introduce rich semantic information from it.
The model architecture is shown in \figref{fig:classification},
which is based on Wide\&Deep \cite{cheng2016wide} framework.
The input is a candidate concept $c$, 
and the output is a score, measuring the probability of $c$ being a good e-commerce concept.
In this paper, we denote a char as a single Chinese or English character,
and a segmented word (or term) is a sequence of several chars such as ``Nike'' or ``牛仔裤 (jeans)''.
We perform Chinese word segmentation for all the input concepts before feeding to the model.

In the Deep side,
there are mainly two components.
Firstly, a char level BiLSTM is used to encode the candidate concept $c$
by feeding the char-level embedding sequence $\{\bi{ch}_1, \bi{ch}_2,...\bi{ch}_n\}$
after simple embedding lookup.
After mean pooling,
we get the concept embedding $\bi{c}_1$.
The other component is knowledge-enhanced module.
The input consists of there parts:
1) pre-trained word embeddings; 2) POS tag \cite{toutanova2003feature} embedding using a lookup table; 3) NER label \cite{finkel2005incorporating} embedding using a lookup table.
After concatenate those three embeddings, 
we obtain the input embedding sequence of candidate concept $c$: 
$\{\bi{w}_1, \bi{w}_2,...\bi{w}_m\}$ ($m<n$).
After going through BiLSTM, we use self attention mechanism \cite{vaswani2017attention} to further encode the mutual influence of each word within the concept and get a sequence output 
$\{\bi{w'}_1, \bi{w'}_2,...\bi{w'}_m\}$.
To introduce external knowledge into the model to do commonsense reasoning on short concepts,
we link each word $w$ to its corresponding Wikipedia article if possible.
For example, ``性感 (sexy)'' can be linked to \url{https://zh.wikipedia.org/wiki/%E6%80%A7%E6%84%9F}
	(\url{https://en.wikipedia.org/wiki/Sexy}).
Then we extract the gloss of each linked Wikipedia article as the external knowledge to enhance the feature representation of concept words.
A gloss is a short document to briefly introduce a word.
We employ Doc2vec \cite{le2014distributed} to encode each extracted gloss for word $\bi{w}_i$ as $\bi{k}_i$.
Then, we get the representation of the knowledge sequence after a self attention layer:
$\{\bi{k'}_1, \bi{k'}_2,...\bi{k'}_m\}$.
We concatenate  $\bi{w'}_i$ as $\bi{k'}_i$ and use max-pooling to get the final knowledge-enhanced representation of candidate concept $\bi{c}_2$.

In the Wide side, 
we mainly adopt pre-calculated features such as the number of characters and words of candidate concept,
the perplexity of candidate concept calculated by 
a BERT \cite{devlin2018bert} model specially trained on e-commerce corpus, and other features like the popularity of each word appearing in e-commerce scenario.
After going through two fully connected layers, 
we get the wide feature representation $\bi{c}_3$.

The final score $\hat y_c$ is calucalated by concatenating the three embedding $\bi{c}_1$, $\bi{c}_2$ and $\bi{c}_3$ then going through a MLP layer.
We use point-wise learning with the negative log-likelihood objective function to learn the parameters of our model:
\begin{equation}
\mathscr{L} = -\sum_{(c)\in D^+}{\log \hat y_c} + \sum_{(c)\in D^-}{\log (1-\hat y_c)}
\end{equation}
where $D^+$ and $D^-$ are the good and bad e-commerce concepts.

We expect this model can help filter out most of bad candidate concepts generated in the first step.
To strictly control the quality, we randomly sample a small portion of every output batch which passes the model checking to ask domain experts to manually annotate.
Only if the accuracy riches a certain threshold,
the whole batch of concepts will be added into AliCoCo.
Besides, the annotated samples will be added to training data to iteratively improve the model performance.


\subsection{Understanding}
\label{sec:tagging}
For those good e-commerce concepts which are directly mined from text corpora,
they are isolated phrases waiting to be integrated into AliCoCo.
To better understand (or interpret) those user needs (aka. e-commerce concepts),
it is a vital step to link them to the layer of primitive concepts.
We call the main task as \textit{``e-commerce concept tagging''}.
Revisit the example shown in \secref{sec:overview}, 
given an surface from ``outdoor barbecue'', 
we need to infer that ``outdoor'' is a ``\textit{Location}'' and ``barbecue'' is an ``\textit{Event}''.
However, word ``barbecue'' can also be a movie in the layer of primitive concepts, so it may be recognized into the class of ``\textit{IP}''.
We formulate this task as a \textit{short text} Named Entity Recognition (NER) problem, which is more challenging to a normal NER task since
concept phrases here are too short (2-3 words on average).
Lack of contextual information make it harder to disambiguate between different classes.

\begin{figure}[th]
	\centering
	\epsfig{file=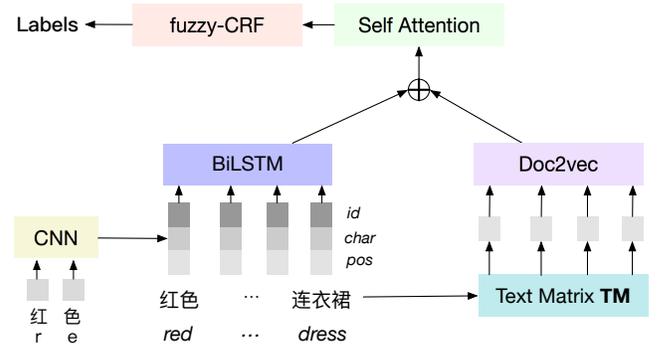, width=\columnwidth}
	\caption{Overview of text-augmented deep NER model for e-commerce concept tagging.
	}
	\label{fig:tagging}
\end{figure}

To overcome the above challenges,
we propose a text-augmented deep NER model with fuzzy CRF,
shown in \figref{fig:tagging}.
The input of this task is a sequence of concept word $\{\bi{w}_1, \bi{w}_2,...\bi{w}_m\}$ after Chinese word segmentation, 
while the output is a sequence of same length $\{\bi{y}_1, \bi{y}_2,...\bi{y}_m\}$ denoting the class labels for each word with In/Out/Begin (\textbf{I/O/B}) scheme.
The model consisting of two components: text-augmented concept encoder and fuzzy CRF layer.

\subsubsection{Text-augmented concept encoder}
To leverage informative features in the representation layer,
we employ word-level, char-level features and position features.
We randomly initialize a lookup table to obtain an embedding for every character. Let $C$ be the vocabulary of characters, a word $w_i$ can be represented as a sequence of character vectors: $\{\bi{c}_1^i, \bi{c}_2^i, ..., \bi{c}_t^i\}$, where $\bi{c}_j^i$ is the vector for the $j$-th character in the word $w_i$ and $t$ is the word length. 
Here we adopt a convolutional neural network (CNN)  architecture to extract the char-level features $\bi{c}_i$ for each word $w_i$. 
Specifically, we use a convolutional layer with window size $k$ to involve the information of neighboring characters for each character.
A max pooling operation is then applied to output the final character representation as follows:
\begin{equation}
	 \bi{c}^i_j = \cnn([\bi{c}_{j-k/2}^i, ..., \bi{c}_j^i, ..., \bi{c}_{j+k/2}^i])
\end{equation}
\begin{equation}
	\bi{c}_i = \maxpool([\bi{c}^i_0, ... \bi{c}^i_j, ...]) 
\end{equation}
To capture word-level features, we use pre-trained word embeddings from GloVe \cite{pennington2014glove} to map a word into a real-valued vector $\bi{x}_i$ , as the initialized word features and will be fine-tuned during training. 
Furthermore,
we calculate part-of-speech tagging features $\bi{p}_i$.
Finally, we obtain the word representation $\bi{w}_i$ by concatenating three embeddings:
\begin{equation}
\bi{w}_i =[\bi{x}_i;\bi{c}_i;\bi{p}_i].
\end{equation}
Similar to the classification model introduced in the previous task,
we feed the sequence of word representations to BiLSTM layer to obtain hidden embeddings $\{\bi{h}_1, \bi{h}_2, ..., \bi{h}_m\}$.
To augment our model with more textual information,
we construct a textual embedding matrix $\textbf{TM}$ by mapping each word back to large-scale text corpus to extract surrounding contexts and encode them via Doc2vec.
Thus, we lookup each word $w_i$ in $\textbf{TM}$ to obtain a text-augmented embedding $\bi{tm}_i$.
We concatenate $\bi{h}_i$ and $\bi{tm}_i$ then use a self attention layer to adjust the representations of each words by considering the augmented textual embeddings of surrounding words, aiming to obtain better feature representations for this task:
\begin{equation}
\bi{h'}_i =\selfatt([\bi{h}_i;\bi{tm}_i]).
\end{equation}

\subsubsection{Fuzzy CRF layer}

\begin{figure}[th]
	\centering
	\epsfig{file=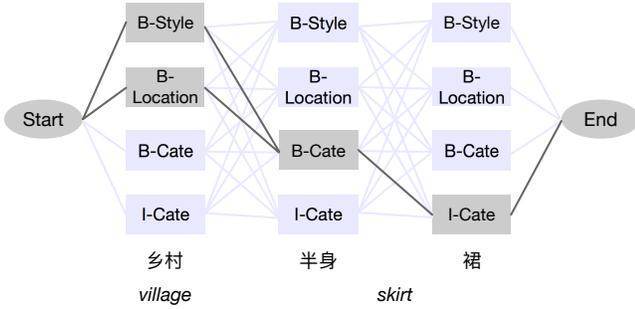, width=\columnwidth}
	\caption{A real example in fuzzy CRF layer.}
	\label{fig:fuzzy}
\end{figure}

Following the concept encoding module, 
we feed the embeddings to a CRF layer.
Different from normal CRF,
we use a fuzzy CRF \cite{shang2018learning} to better handle the disambiguation problem since the valid class label of each word is not unique and this phenomenon is more severe in this task since our concept is too short.
\figref{fig:fuzzy} shows an example, where 
the word ``乡村 (village)'' in the e-commerce concept ``乡村半身裙 (village skirt)'' can linked to the primitive concept ``\textit{空间: 乡村 (Location: Village)}'' or ``\textit{风格: 乡村 (Style: Village)}''.
They both make sense.
Therefore, we adjust the final probability 
as 
\begin{equation}
L(y|\bi{X}) = \frac{\sum_{\hat y \in Y_{possible}}e^{s(X, \hat y)}} {\sum_{\hat y \in Y_{X}}e^{s(X, \hat y)}}.
\end{equation}
where $Y_{X}$ means all the possible label sequences for sequence $X$, and $Y_{possible}$ contains all the possible label sequences.

\section{Item Association}

\label{sec:item}

Items are the most essential nodes in any e-commerce knowledge graph, since the ultimate goal of e-commerce platform is to make sure that customers can easily find items that satisfy their needs.
So far, we conceptualize user needs as e-commerce concepts and interpret
them using the structured primitive concepts.
The last thing is to associate billions of items in Alibaba with all the concepts (both primitive and e-commerce) to form the complete AliCoCo.

Since primitive concepts are similar to single-value tags and properties, 
the mapping between primitive concepts and items are relatively straightforward.
Therefore, in this section, we mainly introduce the methodology of associating items with e-commerce concepts, where the latter ones representing certain shopping scenarios usually carry much more complicated semantic meanings.
Besides, the association between an e-commerce concept and certain items can not be directly inferred from the association between corresponding primitive concepts and their related items due to a phenomenon called ``semantic drift''.
For example, charcoals are necessary when we want to hold an ``outdoor barbecue'',
however, they have nothing to do with primitive concept ``\textit{Location: Outdoor}''.

\begin{figure}[th]
	\centering
	\epsfig{file=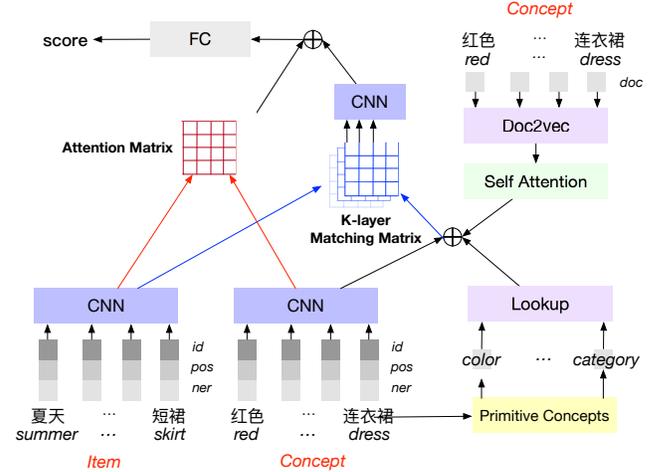, width=1\columnwidth}
	\caption{Overview of knowledge-aware deep semantic matching model for association between e-commerce concepts and items.
	}
	\label{fig:matching}
\end{figure}

We formulate this task as \textit{semantic matching} between texts \cite{huang2013learning,pang2016text,yang2019simple}, since we only use textual features of items at current stage. 
The main challenge to associate e-commerce concepts with related items is that the length of the concept is too short so that limited information can be used.
Due to the same reason, there is a high risk that some of less important words may misguide the matching procedure.
To tackle it, we propose a knowledge-aware deep semantic matching model shown in \figref{fig:matching}.
The inputs are a sequence of concept words and a sequence of words from the title of a candidate item.
We obtain input embeddings concatenating pre-trained word embeddings of two sequences with their POS tag embedding and NER tag embedding (similar to \secref{sec:tagging}):
$\{\bi{w}_1, \bi{w}_2,...\bi{w}_m\}$
and 
$\{\bi{t}_1, \bi{t}_2,...\bi{t}_l\}$.
we adopt wide CNNs with window size $k$ to encode the concept and item respectively:
\begin{equation}
\bi{w'}_i = \cnn([\bi{w}_{i-k/2}, ..., \bi{w}_i, ..., \bi{w}_{i+k/2}])
\end{equation}
\begin{equation}
\bi{t'}_i = \cnn([\bi{t}_{i-k/2}, ..., \bi{t}_i, ..., \bi{t}_{i+k/2}])
\end{equation}
Intuitively, different words in the concept should share different weights when matching to the item, and vice versa.
Therefore, we apply attention mechanism \cite{bahdanau2014neural,luong2015effective} in our model.
An attention matrix is used to model the two-way interactions simultaneously. 
The values of attention matrix are defined as below:
\begin{equation}
att_{i,j} = \bi{v}^T\tanh(\bi{W}_1\bi{w'}_i+\bi{W}_2\bi{t'}_j)
\end{equation}
where $i \in [1, m]$ and $j \in [1, l]$, $\bi{v}$, $\bi{W}_1$ and $\bi{W}_1$ are parameters.
Then the weight of each concept word $w_i$ and title word $t_i$ can be calculated as:
\begin{equation}
	\alpha_{wi} = \frac{exp(\sum_{j}att_{i,j})}{\sum_{i}exp(\sum_{j}att_{i,j})}
\end{equation}
\begin{equation}
\alpha_{tj} = \frac{exp(\sum_{i}att_{i,j})}{\sum_{j}exp(\sum_{i}att_{i,j})}
\end{equation}
Then, we obtain concept embedding \bi{c} as:
\begin{equation}
	\bi{c} = \sum_{i}\alpha_{wi}\bi{w'}_i
\end{equation}
and item embedding \bi{i} similarly.

To introduce more informative knowledge to help semantic matching,
we obtain the same knowledge embedding sequence in \secref{sec:classification}:
\begin{equation}
	\bi{k}_i = \docvec(Gloss(w_i))
\end{equation}
Besides, we obtain class label id embedding $\bi{cls}_j$
of $j$th primitive concept linked with current e-commerce concept.
Thus, there are three sequences on the side of concept:
\begin{eqnarray*}
& \{\bi{kw}_i\} = \{\bi{kw}_1,\bi{kw}_2,...\bi{kw}_{2*m+m'}\} = \\
& \{\bi{w}_1,\bi{w}_2,...\bi{w}_m,\bi{k}_1,\bi{k}_2,...\bi{k}_m,\bi{cls}_1,\bi{cls}_2,...\bi{cls}_{m'}\}
\end{eqnarray*}
where $m'$ is the number of primitive concepts.
In the side of item, we directly use the sequence of word embedding $\{\bi{t}_i\} = \{\bi{t}_1, \bi{t}_2,...\bi{t}_l\}$.
Then, we adopt the idea of Matching Pyramid \cite{pang2016text}, 
the values of matching matrix in $k$th layer are defined as below:
\begin{equation}
	match_{i,j}^k = \bi{kw}_i^{T}\bi{W}_k\bi{t}_j
\end{equation}
where $i \in [1, 2*m+m']$ and $j \in [1, l]$.
Each layer of matching matrix are then fed to 2-layer of CNNs and max-pooling operation to get a matching embedding $\bi{ci}^k$.
The final embedding of matching pyramid $\bi{ci}$ is obtained by:
\begin{equation}
\bi{ci} = \mlp([;\bi{ci}^k;])
\end{equation}

The final score measuring the probability is calculated as:
\begin{equation}
	score = \mlp([\bi{c};\bi{i};\bi{ci}])
\end{equation}


\section{Evaluations}
\label{sec:eval}
In this section, we first give a statistical overview of AliCoCo.
Next we present experimental evaluations for five main technical modules 
during the construction of AliCoCo.

\subsection{Overall Evaluation}

\tabref{tab:data} shows the statistics of AliCoCo.
There are $2,853,276$ primitive concepts and $5,262,063$ e-commerce concepts in total at the time of writing.
There are hundreds of billions of relations in AliCoCo, including $131,968$ isA relations within \textit{Category} in the layer of primitive concepts and $22,287,167$ isA relations in the layer of e-commerce concepts.
For over $3$ billion items in Alibaba, $98\%$ of them are linked to AliCoCo.
Each item is associated with $14$ primitive concepts and $135$ e-commerce concepts on average.
Each e-commerce concept is associated with $74,420$ items on average.
The number of relations between e-commerce concept layer and primitive concept layer is $33,495,112$.

AliCoCo is constructed semi-automatically.
For those nodes and relations mined by models, 
we will randomly sample part of data and ask human annotators to label.
Only if the accuracy achieves certain threshold,
the mined data will be added into AliCoCo to ensure the quality.
Besides, for those dynamic edges (associated with items),
we monitor the data quality regularly.

\begin{table}[th]
	\small
	\begin{tabular}{|l|l|l|l|}
		\hline
		\multicolumn{4}{|l|}{\textbf{Overall}} \\ 
		\hline
		\multicolumn{2}{|l|}{\# Primitive concepts} &
		\multicolumn{2}{l|}{2,853,276}  \\
		\hline
		\multicolumn{2}{|l|}{\# E-commerce concepts} &  
		\multicolumn{2}{l|}{5,262,063}  \\ 
		\hline
		\multicolumn{2}{|l|}{\# Items} &
		\multicolumn{2}{l|}{> 3 billion} \\ 
		\hline
		\multicolumn{2}{|l|}{\# Relations} &
		\multicolumn{2}{l|}{> 400 billion} \\ 
		\hline
		\hline
		\multicolumn{4}{|l|}{\textbf{Primitive concepts}} \\ 
		\hline
		\# Audience &\# Brand &\# Color &\# Design  \\
		\cline{1-4}
		15,168 & 879,311 & 4,396 & 744 \\
		\cline{1-4}
		\# Event & \# Function & \# Category &\# IP  \\
		\cline{1-4}
		 18,400 & 16,379 & 142,755 & 1,491,853
		  \\
		\cline{1-4}
		\# Material & \# Modifier & \# Nature &\# Organization  \\
		\cline{1-4}
		 4,895 & 106 & 75 & 5,766 \\
		\cline{1-4}
		\# Pattern & \# Location & \# Quantity &\# Shape  \\
		\cline{1-4}
		 486 & 267,359 & 1,473 & 110 \\
		\cline{1-4}
		\# Smell & \# Style & \# Taste &\# Time  \\
		\cline{1-4}
		 9,884 & 1,023 & 138 & 365 \\
		\hline
		\hline
		\multicolumn{4}{|l|}{\textbf{Relations}} \\ 
		\hline
		\multicolumn{2}{|l|}{\# IsA in primitive concepts} &
		\multicolumn{2}{l|}{131,968 (only in \textit{Category})}  \\
		\hline
		\multicolumn{2}{|l|}{\# IsA in e-commerce concepts} &
		\multicolumn{2}{l|}{22,287,167}  \\
		\hline
		\multicolumn{2}{|l|}{\# Item - Primitive concepts} &
		\multicolumn{2}{l|}{21 billion}  \\
		\hline
		\multicolumn{2}{|l|}{\# Item - E-commerce concepts} &
		\multicolumn{2}{l|}{405 billion}  \\
		\hline
		\multicolumn{2}{|l|}{\# E-commerce - Primitive cpts} &
		\multicolumn{2}{l|}{33,495,112}  \\
		\hline
	\end{tabular}
	\caption{Statistics of AliCoCo at the time of writing.}
	\label{tab:data}
\end{table}

To evaluate the coverage of actual shopping needs of our customers, we sample $2000$ search queries at random and manually rewrite them into coherent word sequences, then we search in AliCoCo to calculate the coverage of those words. 
We repeat this procedure every day, in order to detect new trends of user needs in time. 
AliCoCo covers over $75\%$ of shopping needs on average in continuous $30$ days, while this number is only $30\%$ for the former ontology mentioned in \secref{sec:intro}.

\subsection{Primitive Concept Mining}
\label{sec:eval_mining}
After defining $20$ different domains in the taxonomy,
we quickly enlarge the size of primitive concepts by introducing knowledges from several existing structured or semi-structured knowledge bases in general-purpose domain.
During this step, vocabulary sizes of domains such as $Location$, $Organization$ and $IntellectulProperty$ can be quickly enlarged.
Other domains are for e-commerce use, and we mainly leverage the existing e-commerce semi-structured data: CPV, since most of $Property$s can be matched to our domains such as $Brand$, $Color$, $Material$, etc.

After rule based alignments and cleaning, 
around $2M$ primitive concepts can be drawn from multiple sources.
We adopt the idea of distant supervision to
generate a large amount of training samples,
in order to mine new concepts.
We use a dynamic programming algorithm of max-matching to match words in the text corpora and then assign each word with its domain label in IOB scheme using existing primitive concepts. We filter out sentences whose matching result is ambiguous and only reserve those that can be perfectly matched (all words can be tagged by only one unique label) as our training data.
We generate around $6M$ training data in this way.
In each epoch of processing $5M$ sentences, 
our mining model is able to discover around $64K$ new candidate concepts on average.
After manually checking the correctness by crowdsourcing services, 
around $10K$ correct concepts can be added into our vocabulary in each round.
The mining procedure is continuously running, 
and the total number of primitive concepts from all $20$ domains 
is $2,758,464$ at the time of writing.

\subsection{Hypernym Discovery}

In order to organize all the primitive concepts into a fine-grained taxonomy,
we propose an active learning framework to iteratively 
discover isA relation between different primitive concepts.
To demonstrate the superior of our framework, 
we perform several experiments on a ground truth dataset collected after the taxonomy is constructed.
We randomly sample 3,000 primitive concepts in the class of ``\textit{Category}'' which have at least one hypernym, and retrieve 7,060 hyponym-hypernym pairs as positive samples.
We split the positive samples into training / validation / testing sets (7:2:1).
The search space of hypernym discovery is actually the whole vocabulary, 
making the number and quality of negative samples very important in this task.
The negative samples of training and validation sets are automatically generated from positive pairs
by replacing the hypernym of each pair with a random primitive concept from ``\textit{Category}'' class.
In the following experiments, mean average precision (MAP), mean reciprocal rank (MRR) and precision at rank 1 (P@1) are used as evaluation metrics. 

\begin{figure}[th]
	\centering
	\epsfig{file=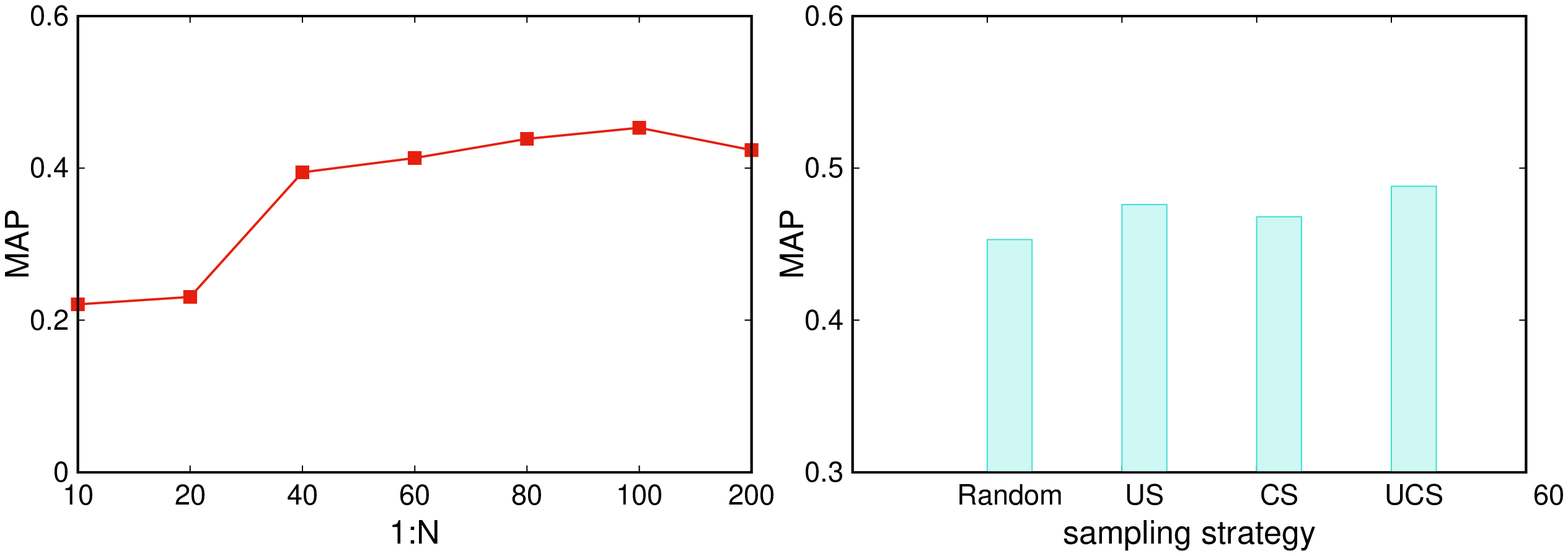, width=\columnwidth}
	\caption{Left: the influence of different negative sample sizes in hypernym discovery on test set. Right: the best performance of different sampling strategies in active learning.}
	\label{fig:isa}
\end{figure}

To verify the appropriate number of negative samples for each hyponym during training,
we perform an experiment shown in \figref{fig:isa}(left), 
where $N$ in x-axis represents the ratio of negative samples over positive samples for each hyponym.
The results indicate different size of negative samples influence the performance differently.
As $N$ gradually increases, the performance improves and achieves best around $100$.
Thus, we construct the candidate training pool in the following active learning experiment with a size of $500,000$.

\tabref{tab:isa} shows experimental results of different sampling strategies during our active learning framework, 
where $Random$ means training using the whole candidate pool without active learning.
We set the select data size $K$ as $25,000$ in each iteration as mentioned in \secref{sec:isa}.
When it achieves similar MAP score in four active learning strategies,
we can find that all the active learning sampling strategies can reduce labeled data to save considerable manual efforts.
UCS is the most economical sampling strategy, which only needs $325k$ samples, reducing $35\%$ samples comparing to random strategy.
It indicates that high confident negative samples are also important in the task of hypernym discovery.

\begin{table}[th]
	\centering
	\begin{tabular}{l|c|c|c|c|c}
		\hline
		Strategy & Labeled Size &  MRR & MAP & P@1 & Reduce \\
		\hline
		Random & 500k & 58.97 & 45.30 & 45.50 & - \\
		US  & 375k &  59.66 & 45.73 & 46.00 & 150k\\
		CS & 400k &  58.96 & 45.22 & 45.30 & 100k\\
		UCS  & 325k &  59.87 & 46.32 & 46.00 & 175k \\
		\hline
	\end{tabular}
	\caption{Experimental results of different sampling strategy in hypernym discovery.}
	\label{tab:isa}
\end{table}

In \figref{fig:isa} (right), 
we show the best performance of each sampling strategies during the whole training procedure.
UCS outperforms other three strategies and achieves a highest MAP of $48.82\%$, showing the importance of selecting the most valuable samples during model training.

\subsection{E-commerce Concept Classification}

In this subsection,
we mainly investigate how each component of our model influences the performance in the task of judging whether a candidate e-commerce concept satisfy the criteria or not (\secref{sec:classification}).

We randomly sample a large portion of e-commerce concepts from the candidate set and ask human annotators to label. The annotation task lasts for several months until we get enough training samples. The final dataset consists of $70k$ samples (positive: negative= 1: 1). Then we split the dataset into 7:1:2 for training, validation and testing. 

\begin{table}[th]
	\centering
	\begin{tabular}{l|c}
		\hline
		Model & Precision   \\
		\hline
		Baseline (LSTM + Self Attention) & 0.870 \\
		+Wide  & 0.900 \\
		+Wide \& BERT & 0.915 \\
		+Wide \& BERT \& Knowledge & \textbf{0.935} \\
		\hline 
	\end{tabular}
	\caption{Experimental results in shopping concept generation. }
	\label{tab:concept}
\end{table}

Results of ablation tests are shown in \tabref{tab:concept}.
Comparing to the baseline, which is a base BiLSTM with self attention architecture, adding wide features such as different syntactic features of concept improves the precision by $3\%$ in absolute value.
When we replace the input embedding with BERT output,
the performance improves another $1.5\%$, 
which shows the advantage of rich semantic information
encoded by BERT.
After introducing external knowledge into our model,
the final performance reaches to $0.935$, improving by a relative gain of $7.5\%$ against the baseline model, indicating that leveraging external knowledge benefits commonsense reasoning on short concepts.

\subsection{E-commerce Concept Tagging}

To associate those e-commerce concepts which are directly mined from text corpus to the layer of primitive concepts,
we propose a text-augmented NER model with fuzzy CRF mentioned in \secref{sec:tagging}
to link an e-commerce concept to its related primitive concepts.
We randomly sample a small set ($7,200$) of e-commerce concepts
and ask human annotators to label the correct class labels for each primitive concepts within the e-commerce concepts.
To enlarge the training data, 
we use the similar idea of distant supervision mentioned in \secref{sec:eval_mining}
to automatically generate $24,000$ pairs of data.
Each pair contains a compound concept and its corresponding gold sequence of domain labels.
We split $7,200$ pairs of manually labeled data into 
$4,800/1,400/1,000$ for training, validation and testing.
$24,000$ pairs of distant supervised data are added into training set to help learn a more robust model.

\begin{table}[th]
	\centering
	\begin{tabular}{l|c|c|c}
		\hline
		Model & Precision &  Recall & F1  \\
		\hline
		Baseline & 0.8573 & 0.8474 &	0.8523 \\
		+Fuzzy CRF  & 0.8731 &  0.8665 & 0.8703 \\
		+Fuzzy CRF \& Knowledge & \textbf{0.8796} &  \textbf{0.8748} & \textbf{0.8772} \\
		\hline
	\end{tabular}
	\caption{Experimental results in shopping concept tagging.}
	\label{tab:linking}
\end{table}

Experimental results are shown in \tabref{tab:linking}.
Comparing to baseline which is a basic sequence labeling model with Bi-LSTM and CRF, 
adding \textit{fuzzy CRF} improves 1.8\% on F1, 
which indicates such multi-path optimization in CRF layer actually contributes to disambiguation.
Equipped with external knowledge embeddings to further enhance the textual information, 
our model continuously outperform to $0.8772$ on F1.
It demonstrates that introducing external knowledge can benefit tasks dealing with short texts with limited contextual information.

\subsection{Concept-Item Semantic Matching}

In this subsection,
we demonstrate the superior of our semantic matching model
for the task of associating e-commerce concepts with billion of items in Alibaba.
We create a dataset with a size of $450m$ samples, among which $250m$ are positive pairs and $200m$ are negative pairs.
The positive pairs comes from strong matching rules and user click logs of the running application on Taobao mentioned in \secref{sec:intro}.
Negative pairs mainly comes from random sampling.
For testing, we randomly sample $400$ e-commerce concepts,
and ask human annotator to label based on a set of candidate pairs.
In total, we collect $200k$ positive pairs and $200k$ negative pairs as testing set.

\begin{table}[th]
	\centering
	\begin{tabular}{l|c|c|c}
		\hline
		Model & AUC & F1 & P@10   \\
		\hline
		BM25 & - & - & 0.7681 \\
		DSSM \cite{huang2013learning}  & 0.7885 & 0.6937 & 0.7971  \\
		MatchPyramid \cite{pang2016text} & 0.8127 & 0.7352 & 0.7813  \\
		RE2 \cite{yang2019simple} \ & 0.8664 & 0.7052 & 0.8977  \\
		\hline
		Ours & 0.8610 & 0.7532 & 0.9015  \\
		Ours + Knowledge & \textbf{0.8713} & \textbf{0.7769} & \textbf{0.9048}  \\
		\hline 
	\end{tabular}
	\caption{Experimental results in semantic matching between e-commerce concepts and items.}
	\label{tab:matching}
\end{table}

\tabref{tab:matching} shows the experimental result, 
where F1 is calculated by setting a threshold of $0.5$.
Our knowledge-aware deep semantic matching model outperforms all the baselines in terms of AUC, F1 and Precision at $10$,
showing the benefits brought by external knowledge.
To further investigate how knowledge helps, 
we dig into cases. Using our base model without knowledge injected,
the matching score of concept ``中秋节礼物 (Gifts for Mid-Autumn Festival)'' and item ``老式大月饼共800g云南特产荞三香大荞饼荞酥散装多口味 (Old big moon cakes 800g Yunnan...)'' is not confident enough to associate those two, since the texts of two sides are not similar.
After we introduce external knowledge for ``中秋节 (Mid-Autumn Festival)'' such as ``中秋节自古便有赏月、吃月饼、赏桂花、饮桂花酒等习俗。(It is a tradition for people to eat moon cakes in Mid-Autumn...)'', 
the attention score for ``中秋节 (Mid-Autumn Festival)'' and ``月饼 (moon cakes)'' increase to bridge the gap of this concept-item pair.

 \section{Applications}
\label{sec:application}
AliCoCo has already supported a series of downstream applications in Alibaba's ecosystem, especially in search and recommendation, two killer applications in e-commerce.
In this section, we introduce some cases we already succeed, those we are attempting now, and some other we would like to try in the future.

\subsection{E-commerce Search}

\subsubsection{Search relevance}

Relevance is the core problem of a search engine, and one of the main challenges is the vocabulary gap between user queries and documents. 
This problem is more severe in e-commerce since language in item titles is more professional. 
Semantic matching is a key technique to bridge the gap in between to improve relevance.
IsA relations is important in semantic matching. 
For example, if a user search for a ``top'', search engine may classify those items whose title only contains ``jacket'' but without ``top'' as irrelevance.
Once we have the prior knowledge that ``jacket is a kind of top'', 
this case can be successfully solved.
Comparing to a former category taxonomy, which only has 15k different category words and 10k isA relations,
AliCoCo containing 10 times categories words and isA relations.
Offline experiments show that our data improves the performance of the semantic matching model by $1\%$ on AUC, 
and online tests show that the number of relevance bad cases is dropped by $4\%$, meaning user satisfaction is improved.

\subsubsection{Semantic search \& question answering}
As shown in \figref{fig:cloud}(a),
semantic search empowered by AliCoCo is ongoing at the time of writing. 
Similar to searching ``China'' on Google and then getting a knowledge card on the page with almost every important information of China, 
we are now designing a more structured way to display the knowledge of ``Tools you need for baking'' once a customer searching for ``baking''.
On the other hand, this application requires a high accuracy and recall of relations, which are still sparse in the current stage of AliCoCo.
Question answering is a way of demonstrating real intelligence of a search engine. 
Customers are used to keyword based search for years in e-commerce.
However, at some point we may want to ask an e-commerce search engine ``What should I prepare for hosting next week's barbecue?''. 
We believe AliCoCo is able to provide ample imagination towards this goal with continuous efforts to integrate more knowledge especially concerning common sense.

\subsection{E-commerce Recommendation}

\subsubsection{Cognitive recommendation}
As we introduce in \secref{sec:intro},
a natural application of e-commerce concepts is directly recommending them to users together with 
its associated items.
In the snapshot shown in \figref{fig:cloud}(b), concept ``Tools for Baking'' is displayed as a card, with the picture of a representative item.
Once users click on this card, 
it jumps to a page full of related items such as egg scrambler and strainer.
We perform thorough offline and online experiments in a previous work
\cite{luo2019conceptualize}.
It has already gone into production for more than 1 year with high click-through rate and satisfied GMV (Gross Merchandise Value).
According to a survey conducted by online users, 
this new form of recommendation brings more novelty and further improve user satisfaction. 
This application is totally based on the complete functionality of AliCoCo, which demonstrates its great value and potential.

\subsubsection{Recommendation reason}

The advantages of e-commerce concepts include its clarity and brevity, which make them perfect recommendation reasons to display when recommending items to customers.
This idea is currently experimented at the time of writing.

\section{Related Work}
\label{sec:related}

Great human efforts have been devoted to construct open domain KGs such as Freebase \cite{bollacker2008freebase} and DBpedia \cite{auer2007dbpedia}, 
which typically describe specific facts with well-defined type systems rather than inconsistent concepts from natural language texts. 
Probase \cite{wu2012probase} constructs a large-scale probabilistic taxonomy of concepts, organizing general concepts using isA relations. 
Different from AliCoCo, concepts in Probase do not have classes so that semantic heterogeneity is handled implicitly.
From this perspective, the structure of AliCoCo is actually more similar to KGs with a type system such as Freebase.
ConceptNet \cite{speer2012representing} tries to include common sense knowledge by recognizing informal relations between concepts, where the concepts could be the conceptualization of any human knowledge such as ``games with a purpose'' appearing in free texts.
Inspired by the construction of open-domain KGs, different kinds of KGs in e-commerce are constructed
to describe relations among users, items and item attributes \cite{catherine2017explainable,ai2018learning}.
One famous example is the ``Product Knowledge Graph'' (PG) \footnote{\url{http://conferences.cis.umac.mo/icde2019/wp-content/uploads/2019/06/icde-2019-keynote-luna-dong.pdf}} of Amazon, another e-commerce giant in the world. The major difference is that they do not focus on user needs as we do.
In AliCoCo, we formally define user needs and introduce a new type of nodes named e-commerce concepts to explicitly represent various shopping needs and further link them to the layer of primitive concepts for semantic understanding. 
Although we do not discuss much, AliCoCo can be connected to open-domain KGs through the layer of primitive concepts (e.g. IP, Organization, etc) just like PG, 
making it more powerful.

\section{Conclusion}
\label{sec:conclusion}

In this paper,
we point out that there is a huge semantic gap between user needs and current ontologies in most e-commerce platforms.
This gap inevitably leads to a situation where e-commerce search engine and recommender system can not understand user needs well, 
which, however, are precisely the ultimate goal of e-commerce platforms try to satisfy.
To tackle it, we introduce a specially designed e-commerce cognitive concept net ``AliCoCo'' practiced in Alibaba, trying to conceptualize user needs as various shopping scenarios, 
also known as ``e-commerce concepts''. 
We present the detailed structure of AliCoCo and introduce how it is constructed with abundant evaluations.
AliCoCo has already benefited a series of downstream e-commerce applications in Alibaba.
Towards a subsequent version, our future work includes: 
1) Complete AliCoCo by mining more unseen relations containing commonsense knowledge, for example,
``boy's T-shirts'' implies the ``\textit{Time}'' should be ``\textit{Summer}'', even though term ``summer'' does not appear in the concept.
2) Bring probabilities to relations between concepts and items.
3) Benefit more applications in e-commerce or even beyond e-commerce.

%
%

\section{acknowledgment}
We deeply thank Mengtao Xu, Yujing Yuan, Xiaoze Liu, Jun Tan and Muhua Zhu for their efforts on the construction of AliCoCo.

\bibliographystyle{ACM-Reference-Format}
\bibliography{ref}
\end{CJK}
\end{document}